\documentstyle[12pt,a4]{article}

\begin{document}
\thispagestyle{empty}
\vspace*{-1.5cm}
\hfill {\small KL--TH 95/22} \\[8mm]

\message{reelletc.tex (Version 1.0): Befehle zur Darstellung |R  |N, Aufruf
z.B. \string\bbbr}
%
%
\message{reelletc.tex (Version 1.0): Befehle zur Darstellung |R  |N, Aufruf
z.B. \string\bbbr}
%
%
%
%
%
\font \smallescriptscriptfont = cmr5
\font \smallescriptfont       = cmr5 at 7pt
\font \smalletextfont         = cmr5 at 10pt
\font \tensans                = cmss10
\font \fivesans               = cmss10 at 5pt
\font \sixsans                = cmss10 at 6pt
\font \sevensans              = cmss10 at 7pt
\font \ninesans               = cmss10 at 9pt
\newfam\sansfam
\textfont\sansfam=\tensans\scriptfont\sansfam=\sevensans
\scriptscriptfont\sansfam=\fivesans
\def\sans{\fam\sansfam\tensans}
\def\bbbr{{\rm I\!R}} 
\def\bbbn{{\rm I\!N}} 
\def\bbbE{{\rm I\!E}} 
\def\bbbm{{\rm I\!M}}
\def\bbbh{{\rm I\!H}}
\def\bbbk{{\rm I\!K}}
\def\bbbd{{\rm I\!D}}
\def\bbbp{{\rm I\!P}}
\def\bbbone{{\mathchoice {\rm 1\mskip-4mu l} {\rm 1\mskip-4mu l}
{\rm 1\mskip-4.5mu l} {\rm 1\mskip-5mu l}}}
\def\bbbc{{\mathchoice {\setbox0=\hbox{$\displaystyle\rm C$}\hbox{\hbox
to0pt{\kern0.4\wd0\vrule height0.9\ht0\hss}\box0}}
{\setbox0=\hbox{$\textstyle\rm C$}\hbox{\hbox
to0pt{\kern0.4\wd0\vrule height0.9\ht0\hss}\box0}}
{\setbox0=\hbox{$\scriptstyle\rm C$}\hbox{\hbox
to0pt{\kern0.4\wd0\vrule height0.9\ht0\hss}\box0}}
{\setbox0=\hbox{$\scriptscriptstyle\rm C$}\hbox{\hbox
to0pt{\kern0.4\wd0\vrule height0.9\ht0\hss}\box0}}}}

\def\bbbe{{\mathchoice {\setbox0=\hbox{\smalletextfont e}\hbox{\raise
0.1\ht0\hbox to0pt{\kern0.4\wd0\vrule width0.3pt height0.7\ht0\hss}\box0}}
{\setbox0=\hbox{\smalletextfont e}\hbox{\raise
0.1\ht0\hbox to0pt{\kern0.4\wd0\vrule width0.3pt height0.7\ht0\hss}\box0}}
{\setbox0=\hbox{\smallescriptfont e}\hbox{\raise
0.1\ht0\hbox to0pt{\kern0.5\wd0\vrule width0.2pt height0.7\ht0\hss}\box0}}
{\setbox0=\hbox{\smallescriptscriptfont e}\hbox{\raise
0.1\ht0\hbox to0pt{\kern0.4\wd0\vrule width0.2pt height0.7\ht0\hss}\box0}}}}

\def\bbbq{{\mathchoice {\setbox0=\hbox{$\displaystyle\rm Q$}\hbox{\raise
0.15\ht0\hbox to0pt{\kern0.4\wd0\vrule height0.8\ht0\hss}\box0}}
{\setbox0=\hbox{$\textstyle\rm Q$}\hbox{\raise
0.15\ht0\hbox to0pt{\kern0.4\wd0\vrule height0.8\ht0\hss}\box0}}
{\setbox0=\hbox{$\scriptstyle\rm Q$}\hbox{\raise
0.15\ht0\hbox to0pt{\kern0.4\wd0\vrule height0.7\ht0\hss}\box0}}
{\setbox0=\hbox{$\scriptscriptstyle\rm Q$}\hbox{\raise
0.15\ht0\hbox to0pt{\kern0.4\wd0\vrule height0.7\ht0\hss}\box0}}}}

\def\bbbt{{\mathchoice {\setbox0=\hbox{$\displaystyle\rm
T$}\hbox{\hbox to0pt{\kern0.3\wd0\vrule height0.9\ht0\hss}\box0}}
{\setbox0=\hbox{$\textstyle\rm T$}\hbox{\hbox
to0pt{\kern0.3\wd0\vrule height0.9\ht0\hss}\box0}}
{\setbox0=\hbox{$\scriptstyle\rm T$}\hbox{\hbox
to0pt{\kern0.3\wd0\vrule height0.9\ht0\hss}\box0}}
{\setbox0=\hbox{$\scriptscriptstyle\rm T$}\hbox{\hbox
to0pt{\kern0.3\wd0\vrule height0.9\ht0\hss}\box0}}}}

\def\bbbs{{\mathchoice
{\setbox0=\hbox{$\displaystyle     \rm S$}\hbox{\raise0.5\ht0\hbox
to0pt{\kern0.35\wd0\vrule height0.45\ht0\hss}\hbox
to0pt{\kern0.55\wd0\vrule height0.5\ht0\hss}\box0}}
{\setbox0=\hbox{$\textstyle        \rm S$}\hbox{\raise0.5\ht0\hbox
to0pt{\kern0.35\wd0\vrule height0.45\ht0\hss}\hbox
to0pt{\kern0.55\wd0\vrule height0.5\ht0\hss}\box0}}
{\setbox0=\hbox{$\scriptstyle      \rm S$}\hbox{\raise0.5\ht0\hbox
to0pt{\kern0.35\wd0\vrule height0.45\ht0\hss}\raise0.05\ht0\hbox
to0pt{\kern0.5\wd0\vrule height0.45\ht0\hss}\box0}}
{\setbox0=\hbox{$\scriptscriptstyle\rm S$}\hbox{\raise0.5\ht0\hbox
to0pt{\kern0.4\wd0\vrule height0.45\ht0\hss}\raise0.05\ht0\hbox
to0pt{\kern0.55\wd0\vrule height0.45\ht0\hss}\box0}}}}

\def\bbbz{{\mathchoice {\hbox{$\sans\textstyle Z\kern-0.4em Z$}}
{\hbox{$\sans\textstyle Z\kern-0.4em Z$}}
{\hbox{$\sans\scriptstyle Z\kern-0.3em Z$}}
{\hbox{$\sans\scriptscriptstyle Z\kern-0.2em Z$}}}}

\begin{center}
{\Large\bf Sigma models with $A_k$ singularities \\in Euclidean spacetime
of dimension $0 \le D < 4$ and in the limit $N \rightarrow \infty$}
\\[14mm]
{\large J. Maeder and W. R\"uhl}\\
[1cm]
Fachbereich Physik \\
Universit\"at Kaiserslautern\\
P.\ O.\ Box 3049\\
D--67653 Kaiserslautern,
Germany \\[10mm]

{\bf Abstract}
\end{center}

\noindent For the case of the single-O($N$)-vector linear sigma models the
critical
behaviour following from any $A_k$ singularity in the action is worked out in
the
double scaling limit $N \rightarrow \infty$, $f_r \rightarrow f_r^c$, $2 \leq r
\leq k$.
After an exact elimination of Gaussian degrees of freedom, the critical objects
such
as coupling constants, indices and susceptibility matrix are derived for all
$A_k$ and
spacetime dimensions $0 \leq D < 4$. There appear exceptional spacetime
dimensions
where the degree $k$ of the singularity $A_k$ is more strongly constrained than
by
the renormalizability requirement.

\newpage

\section{Introduction}
\indent
Matrix models can be reformulated as representing stochastic triangulated
surfaces
and are thus interpreted as quantum gravity theories. They are treated in
the "double scaling limit" $N \rightarrow \infty , g \rightarrow g_c$
\cite{1}-\cite{3}.
$O(N)$ vector sigma models can be understood in a similar manner as describing
statistical ensembles of branched polymers \cite{4}-\cite{7}. These models can
also be
submitted to a double scaling limit. This is done with $N$-renormalization
group
techniques based on exact recursion relations in $N$ \cite{4}-\cite{10}.

Instead we propose to start from saddle point integrals. Partition functions
are then to leading order represented in the form of generalized Airy functions
($D = 0$, see (\ref{63})) or as a partition function with a new field theoretic
action
($D > 0$, see (\ref{91})). We shall refer to the function (respectively:
functional) in the
exponential as "Airy action" (respectively: "Airy field action").

The saddle point integrals arise from singularities in the original action
when the limit $N \rightarrow \infty$ is performed. Such singularities can
be classified \cite{11,12} and form $s$-dimensional families. Each familiy
possesses $s$ moduli as continuous parameters. If $s = 0$, the families are
discrete and are grouped by their symmetry into A, D and E series. The
A-series can be realized in single-vector sigma models and is the object of
our interest in this article. It has been shown \cite{13,14} that D and E
series
can be realized by two-vector sigma models. In the field theoretic literature
only A-series singularities have been identified before us (the triple scaling
cases (i) and (ii) in \cite{15} are separable as $A_1 \times A_2$, respectively
$A_1 \times A_3$).

By application of diffeomorphisms a singularity can be brought to canonical
form. This canonical form contains the full information defining a
"universality
class of multicritical behaviour" ($A_k$: k-critical). Our aim in this article
is to extract universal quantities such as critical indices
and the universal part of the beta functions for the whole A-series
in dimensions $0 \le D < 4$. For $D > 0$ two kinds of boundary conditions
are considered: finite cube periodic boundary conditions and infinite volume.
The spacetime dimension $D$ is interpolated whenever possible. We obtain
closed algebraic expressions in each case (no infinite sums or integrals).

Our treatment of $D \geq 2$ field theories is restricted to "naive double
scaling",
i.e. renormalizations are neglected. These imply logarithmic modifications,
namely
$N^{\sigma}$ is multiplied with a polynomial in $\log N$ \cite{8,9}. We expect
that
these modifications can also be calculated explicitly (i.e. in terms of
algebraic
expressions) for all cases $A_k$. Some preliminary discussions are presented in
\cite{9} (e.g. introduction of counter terms). These logarithmic modifications
are
also of interest for mathematics, they go beyond the concepts of Arnold's
school.

The authors of \cite{7,9} treated the cases $A_k, \; k \geq 3$, incorrectly.
They
eliminated Gaussian degrees of freedom connected with nonvanishing eigenvalues
of
the Hessian by orthogonal projection along the zero mode eigenvectors. The
orthogonal
structure is produced by the Hessian itself, which looses its meaning at higher
orders
of the Taylor expansion. In fact, it is not difficult to see that additional
"curvature terms" arise first at fourth order ($k=3$). The correct method is
explained
in the text. It is based on the "splitting lemma" (\cite{12}, Theorem 4.5)
which is
proved by the implicit function theorem. Gaussian degrees of freedom, which
each
belong to an $A_1$ singularity, have to be integrated out before the relevant
singularity is isolated.

The Airy functions are central and unambiguously derived in our approach. In
the
$N$-renormalization group approach a constrained Airy function depending on
only
one variable is obtained by solving a differential equation. Namely, let the
sum
in (\ref{63}) run over $1 \leq n \leq k-1$ and set
\[
\zeta = -\zeta_1, \; \zeta_2 = \zeta_3 = \cdots = \zeta_{k-1} = 0, \; \epsilon
= +1
\]
then the resulting function $Y(\zeta)$ satisfies
\[
\left( \zeta- \left( \frac{d}{d\zeta} \right)^{k} \right) Y(\zeta) = 0.
\]
In the same way one can derive a system of differential equations for the
general
case \cite{13,14}.

Asymptotic expansions for large $\zeta$ play an important role in the
$N$-renormalization
group approach \cite{7,10,15}. For the singularity $A_2$ one can use the known
expansions of the standard Airy functions Ai and Bi \cite{16}. For the
generalized
Airy functions of the singularities $A_k, \; k \geq 3$, different orderings of
the
large arguments $\{ \zeta_l \}$ are possible, each one by repeated application
of
saddle point integrations along a chain of reductions
\[
A_k \rightarrow A_{k-1} \rightarrow \cdots \rightarrow A_2 \rightarrow A_1
\]
implying a different asymptotic expansion. Degenerate reduction chains with
intermediate singularities skipped can also occur.

In Section 2 the single-O($N$)-vector sigma model is formulated and transformed
into an effective field theory of two scalar fields. The Hessian of this
effective field theory is the basis for the discussion of singularities. Its
corank must be one in order to admit singularities of type $\{A_k\}^{\infty}
_2$. It was shown in \cite{14} that sigma models containing $r$ O($N$)-vectors
can be formulated such that the Hessian is of corank $r$.

The elimination of the Gaussian degrees of freedom which are all degrees
except one for $\{A_k\}^{\infty}_2$, is a major algebraic issue. It is
formulated and solved in Section 3 for finite volume. Moreover we calculate
critical coupling constants and the position of the saddle point ($r_0$ or
b(0)). Both these results can be carried over to the infinite volume case
(Section 4).

The deviations of the coupling constants from their critical value map to
lowest order linearly on the deformation space. This linear map is
denoted as "susceptibility matrix". Its calculation is the major topic of
Section 4. It can also be inverted explicitly. The double scaling limit
is defined as the combined limit when $N$ tends to infinity and the coupling
constants approach their critical values. In detail it involves the
susceptibility matrix and the critical indices. Both enter also the linear
terms of the beta functions which can thus be given explicitly for all
$\{A_k\}^{\infty}_2$.

In Section 5 we study the case when the sigma model is carried by the whole
of $D < 2$ dimensional space time. The momenta form a continuous spectrum.
A momentum scale $\Lambda$ which is tied to the renormalized mass
of the theory separates small momenta $\{| p | < \Lambda \}$ from large
momenta $\{| p | > \Lambda \}$. The latter belong to Gaussian degrees
of freedom that are integrated out, whereas the former are additional
deformation parameters which under double scaling induce the kinetic energy
term in the Airy field action. The critical indices are modified but the
susceptibility matrix remains unchanged as compared with Section 4.

If the sigma model is carried by infinite spacetime of dimensions $2 \le
D < 4$, the double scaling limit can be performed provided the field theory
resulting from the saddle point integral is renormalizable. Counter terms
have to be introduced \cite{17,9} in the course of the limit and the quantity
$N \Lambda$
is the ultraviolet cutoff. This is studied in Section 6. If we use dimensional
regularization for $2 < D < 4$, all critical objects can be shown to be
analytic continuations of the corresponding quantities for $0 < D < 2$.
However, for $D = 2$ we use a subtraction scheme depending on a mass parameter
$\mu ^2$ and all critical quantities are recalculated.

The type $A_k$ of the
singularity is restricted by renormalizability to
\[
k \le \frac{D+2}{D-2}.
\]
Surprisingly we observe that for "exceptional dimensions"
\[
D_n = 2 \frac{n}{n-1}, \quad n \in \{ 3,4,5,... \}
\]
$k$ is further restricted by
\[
k \le \left\{ \begin{array}{l}
n-1, \mbox{n odd,} \\
n-2, \mbox{n even.} \end{array} \right.
\]
This result is a consequence of the analytic form found for the critical
quantities.

In Section 7 we return to the unstable field theories resulting from saddle
point integrals. Though we make suggestions of how to ascribe a meaning to
them, their properties are unclear. Nevertheless, realistic systems of
statistical mechanics may be described by them and it would be wrong to
neglect them.

\section{The model}
We study conventional sigma models with the action
\begin{equation}
S = \int d^{D}x \left [ \frac12 (\partial _\nu \vec{S}) (\partial _\nu
\vec{S}) + \frac12 \beta^2 \vec{S}^2 + U(\vec{S}^2) \right ]
\label{1}
\end{equation}
$(\vec{S} \in {\bbbr}_N)$

\noindent with the potential
\begin{equation}
U(\sigma) = \sum ^{\infty} _{r=2} \frac{f_r}{r} \sigma^r .
\label{2}
\end{equation}
We shall interpolate the dimension $D$. For the purpose of our investigation it
is not
relevant whether the series (\ref{2}) is finite, analytic or formal.

By a standard functional Fourier transformation and performing some of the
functional integrations we can transform the action (\ref{1}) into an
effective action
\begin{eqnarray}
S_{\mbox{\scriptsize eff}} & = &\int d^Dx \left [ U(\sigma (x)) - i \rho (x)
\sigma (x) \right ]\nonumber \\
& &+ \frac12 \mbox{Tr} \log [- \Delta + \beta^2 + 2i\rho]
\label{3}
\end{eqnarray}
with partition function
\begin{equation}
Z = \int D\sigma D\rho \exp [- N S_{\mbox{\scriptsize eff}} (\sigma ,\rho)].
\label{4}
\end{equation}
This partition function is to be evaluated in the limit $N \rightarrow \infty$.
Application of singularity theory amounts to evaluation of (\ref{4}) by saddle
point integrals.

The system may either be considered on unbounded spacetime or on a cube with
volume $V = L^D$ and periodic boundary conditions. Fourier transforms are
defined by
\begin{equation}
\hat{\alpha}(p) = \int d^Dx \, e^{-ipx} \alpha (x)
\label{5}
\end{equation}
in either case, but the inverse transformations involve either integrations
\[
\int\frac{d^Dp}{(2\pi )^D}
\]
or summations
\[
\frac{1}{V} \sum_{{\mbox{\scriptsize{$p$ from one}}} \atop  {\mbox{\scriptsize
Brillouine zone}}}.
\]
In explicit formulas we will always write integrals.

Let us assume that the saddle point of (\ref{4})
\[
(\sigma_0,\rho_0)
\]
is constant over spacetime. The saddle point is then determined by
\begin{equation}
U^{\prime}(\sigma_0) = i\rho_0
\label{6}
\end{equation}
\begin{equation}
\sigma_0 = \int \frac{d^Dp}{(2\pi )^D} (p^2 + m^2)^{-1}
\label{7}
\end{equation}
where
\begin{equation}
m^2 = \beta^2 + 2i\rho_0
\label{8}
\end{equation}
is assumed positive. In order to render the integral (\ref{7}) convergent,
we limit $D$ to $0 \le D < 2$. Only in the sixth section we will abandon this
constraint.

The fields $\sigma$ and $\rho$ fluctuate
\begin{equation}
\sigma(x) = \sigma_0 (1 + \alpha(x))
\label{9}
\end{equation}
\begin{equation}
\rho(x) = \rho_0 (1 + \beta(x))
\label{10}
\end{equation}
and the n'th order term of $S_{\mbox{\scriptsize eff}}$ in the fluctuations is
denoted
$S^{(n)}_{\mbox{\scriptsize eff}}$. Then
\begin{eqnarray}
S^{(2)}_{\mbox{\scriptsize eff}} = \frac12 \int \frac{d^Dp}{(2\pi )^D}
(\hat{\alpha}(-p),
\hat{\beta}(-p))  \nonumber \\
\left( \begin{array}{cc}
\sigma^2_0 U^{\prime\prime}(\sigma_0) & \sigma_0 r_0 \\
\sigma_0 r_0 & -2r^2_0 \Sigma (p)
\end{array}\right)
\left( \begin{array}{cc}
\hat{\alpha}(p)\\
\hat{\beta}(p)
\end{array}\right)
\label{11}
\end{eqnarray}
where the real quantity
\begin{equation}
r_0 = -i \rho_0
\label{12}
\end{equation}
has been introduced and
\begin{equation}
\Sigma(p) = \int \frac{d^Dq}{(2\pi )^D} [(q^2 + m^2)((q - p)^2 + m^2)]^{-1}.
\label{13}
\end{equation}
The n'th order term $S^{(n)}_{\mbox{\scriptsize eff}}$ is in coordinate space
integrals
\begin{eqnarray}
S^{(n)}_{\mbox{\scriptsize eff}} = \frac{\sigma^n_0 U^{(n)}(\sigma_0)}{n!} \int
d^Dx \alpha(x)^n
\nonumber \\
- \frac{1}{2n} (2r_0)^n \mbox{Tr} \left[(- \Delta+m^2)^{-1} \beta(x) \right]^n.
\label{14}
\end{eqnarray}

The Hessian $S^{(2)}_{\mbox{\scriptsize eff}}$ is diagonalized by
\begin{equation} \left(\begin{array}{cc}
\hat{\alpha}(p)\\
\hat{\beta}(p)
\end{array}\right) =
\left(\begin{array}{cc}
a(p)\\
1 \end{array}\right)
\hat{\xi}(p) +
\left(\begin{array}{cc}
b(p)\\
1 \end{array}\right)
\hat{\eta}(p)
\label{15}
\end{equation}
where the two eigenvectors are orthogonal implying
\begin{equation}
a(p)b(p) = -1
\label{16}
\end{equation}
and have norms squared
\begin{eqnarray}
N_+(p) = a(p)^2 + 1 \nonumber \\
N_-(p) = b(p)^2 + 1   \label{17}.
\end{eqnarray}
Then the Hessian assumes the form
\begin{eqnarray}
S^{(2)}_{{\mbox{\scriptsize eff}}} =  \frac12 \int \frac{d^Dp}{(2\pi )^D}
\big[\lambda_+(p)
N_+(p) \hat{\xi}(-p) \hat{\xi}(p) \nonumber \\
\quad \quad + \lambda_-(p) N_-(p) \hat{\eta}(-p)\hat{\eta}(p) \big]
\label{18}
\end{eqnarray}
with eigenvalues
\begin{eqnarray}
\lambda_{\pm}(p) = \frac12 \left\{ \sigma^2_0 U^{\prime\prime}(\sigma_0)
- 2r^2_0 \Sigma(p) \right. \nonumber \\
\left. \quad \mp \left[(\sigma^2_0 U^{\prime\prime}(\sigma_0) + 2r^2_0
\Sigma(p))^2 + 4 \sigma^2_0 r^2_0 \right] ^{\frac12} \right\}.
\label{19}
\end{eqnarray}

The $p$-dependence of $N_{\pm}(p), a(p), b(p), \lambda_{\pm}(p)$ originates
from $\Sigma(p)$. Since
\begin{equation}
\Sigma(p) = \Sigma(-p)
\label{20}
\end{equation}
all these quantities have the same symmetry. In the infinite volume
case $\Sigma(p)$ depends only on $p^2$ and decreases monotonously from $p^2=0$
to $p^2= \infty$ where it vanishes.

We will discuss only the case that the Hessian becomes singular at $p=0$.
In this case we must have
\begin{equation}
2 \Sigma(0) U^{\prime\prime} (\sigma_0) + 1 = 0
\label{21}
\end{equation}
implying
\begin{equation}
\lambda_-(0) = 0
\label{22}
\end{equation}

\begin{equation}
\lambda_+(0) < 0.
\label{23}
\end{equation}
{}From (\ref{19}) follows moreover

\begin{equation}
\lambda_-(p) > 0, \quad p \neq 0.
\label{24}
\end{equation}

\section{The critical behaviour for fixed finite volume}

The critical behaviour of a singularity of type $A_k$ is
produced by the limit $N \rightarrow \infty$ and is essentially
independent of whether the volume is finite or infinite. Of course the
finite size leads to finite size corrections which we shall neglect here.
If the volume is finite, the momentum spectrum is discrete and the corank
of the Hessian is one. The possible singularities are of type $A_k, k \in
\bbbn$.
The fluctuations
\begin{eqnarray*}
\hat{\xi}(p), \mbox{ all } p \\
\hat{\eta}(p), \mbox{ all }p \neq 0
\end{eqnarray*}
are Gaussian and must be integrated out first. The saddle point around
which the Gaussian integration is performed is fixed by
\begin{equation}
\frac{\partial S_{\mbox{\scriptsize eff}}}{\partial \hat{\xi}(p)} = 0,
\mbox{ all }p
\label{25}
\end{equation}
\begin{equation}
\frac{\partial S_{\mbox{\scriptsize eff}}}{\partial \hat{\eta}(p)} = 0,
\mbox{ all }p \neq 0
\label{26}
\end{equation}
so that its location depends on $\hat{\eta}(0)$. Because of translational
invariance these conditions imply
\begin{equation}
\hat{\xi}(p) = \hat{\eta}(p) = 0, \quad p \neq 0.
\label{27}
\end{equation}
Inserting this into $S_{\mbox{\scriptsize eff}}$ gives a new action
\begin{equation}
\left.\tilde{S}_{\mbox{\scriptsize red}}(\xi_0,\eta_0) = S_{\mbox{\scriptsize
eff}}(\hat{\xi},\hat{\eta})\right|_
{\hat{\xi}(p) = \hat{\eta}(p) = 0 \atop{(p \neq 0)}}
\label{28}
\end{equation}
where we introduced the new variables
\begin{equation}
\xi_0 = \frac{\hat{\xi}(0)}{V}, \eta_0 = \frac{\hat{\eta}(0)}{V}.
\label{29}
\end{equation}

{}From (\ref{25}) remains
\begin{equation}
\frac{\partial \tilde{S}_{\mbox{\scriptsize red}}}{\partial\xi_0}
(\xi_0,\eta_0) = 0
\label{30}
\end{equation}
or explicitly
\begin{equation}
- \lambda_+(0) N_+(0) \xi_0 = \sum^{\infty}_{n=3} \frac{\partial}
{\partial\xi_0} \tilde{S}^{(n)}_{\mbox{\scriptsize red}} (\xi_0,\eta_0).
\label{31}
\end{equation}
A splitting lemma \cite{12} asserts that this elimination equation
can be solved iteratively
\begin{equation}
\xi_0 = H(\eta_0) = \sum^{\infty}_{n=2} a_n\eta^n_0
\label{32}
\end{equation}
and that $H$ exists in a neighborhood of zero as a function. An explicit
determination of the coefficients $\{a_n\}$ at the critical point (the
$\{a_n\}$ are functions of $\{f_r \}$ from (\ref{2}) which assume critical
values $\{f^c_r\}$) is crucial for any further explicit determination of
critical quantities.

Inserting (\ref{32}) into $\tilde{S}_{\mbox{\scriptsize red}}$ (\ref{28}) gives
the reduced
action (neglecting a constant)
\begin{eqnarray}
S_{\mbox{\scriptsize red}}(\eta_0) & = & \sum^{\infty}_{n=2} \frac{g_n}{n}
\eta^n_0 \nonumber \\
& = & \tilde{S}_{\mbox{\scriptsize red}} (H(\eta_0),\eta_0).
\label{33}
\end{eqnarray}
A singularity $A_k$ arises if
\begin{eqnarray}
g_n = 0, \quad 2 \le n \le k \nonumber \\
g_{k+1} \neq 0.
\label{34}
\end{eqnarray}

Now we start the explicit calculation of the critical quantities. Some
technicalities are unavoidable in this context. The function $\Sigma(p)$
(\ref{13}) can be expanded in a power series of $p^2$ (convergence radius
$4m^2$, only the first terms are needed)
\begin{equation}
\Sigma(p) = \sum^{\infty}_{n=0} B_n(p^2)^n.
\label{35}
\end{equation}
 Moreover, we use the standard integrals
\begin{eqnarray}
\Pi_n & = & \int \frac{d^Dp}{(2\pi )^D} (p^2+m^2)^{-n} \nonumber \\
& = & (4\pi)^{-\mu} \frac{\Gamma(n-\mu)}{\Gamma(n)} (m^2)^{\mu-n}
\label{36}
\end{eqnarray}
\[
(\mu = \mbox{$\frac12$} D) .
\]
Then $B_n$ can be expressed in terms of $\Pi_{n+2}$ by
\begin{equation}
B_n = \frac{(-1)^n n!(n+1)!}{(2n+1)!} \Pi_{n+2}.
\label{37}
\end{equation}
We can use $\Pi_1$ and $\Pi_2$ to express all fractional powers of momentum
dimensions, e.g.
\begin{equation}
\Pi_{n+2} = \frac{\Pi_2^{n+1}}{\Pi^n_1} \delta_n
\label{38}
\end{equation}
so that $\delta_n$ is a function of $D$ only (or $\mu = \frac12 D$)
\begin{equation}
\delta_n = \frac{(2-\mu)_n}{(n+1)!(1-\mu)^n}.
\label{39}
\end{equation}

Noting that $\sigma_0 = \Pi_1$ by (\ref{7}) and (\ref{36}), we
normalize the derivation of the potential at the critical point by
\begin{equation}
\left. \Pi^n_1 U^{(n)} (\Pi_1)\right|_{\mbox{\scriptsize crit}} =
\frac{\Pi^2_1}{2\Pi_2} v_n.
\label{40}
\end{equation}
Then the following result can be derived
\begin{equation}
v_{n+2} = \sum_{\mbox{{\scriptsize partitions of $n$}}} (-1)^{\ell -1}
(n+\ell)! \, \prod^{\infty}_{j=1} \frac{\delta^{n_j}}{n_j!}
\label{41}
\end{equation}
\quad $(n \ge 0, v_2 = -1$ from (\ref{21}))

\noindent where $\ell$ is the length of the partition and $n_j$ is the
repetition number of $j$
\begin{equation}
n = \sum^{\infty}_{j=1} jn_j
\label{42}
\end{equation}
\begin{equation}
\ell = \sum^{\infty}_{j=1} n_j.
\label{43}
\end{equation}

This formula has been checked by computer up to $n = 8$. Inserting $\delta_n$
(\ref{39}) into (\ref{41}) gives the simple expression
\begin{equation}
v_{n+2} = (-1)^{n+1} \left( \frac{2-\mu}{1-\mu}\right)_n.
\label{44}
\end{equation}

In the case in which the volume is finite, the definitions
(\ref{38}),(\ref{40})
and the purely algebraic result (\ref{41}) remain valid. But $\delta_n$
(\ref{39})
obtains a finite size correction which is neglected in (\ref{44}). We shall
neglect such corrections also in the sequel.

The next issue is to calculate the critical coupling constants $\{f^c_r\}$
from all $v_n$. For a singularity $A_k$ we normalize the coupling constants
to
\begin{equation}
A_k : f_r = 0 \mbox{ for } r > k.
\label{45}
\end{equation}
In analogy with (\ref{40}) we define
\begin{equation}
\Pi_1^r f^c_r = \frac{\Pi^2_1}{2\Pi_2} p^{(k)}_r(\mu).
\label{46}
\end{equation}
Inverting the system of equations $(n \in \{1,2,...,k-1\})$
\begin{equation}
v_{n+1} = (-1)^n \sum^k_{r=2} (1-r)_n p_r^{(k)}(\mu)
\label{47}
\end{equation}
gives $(r \in \{2,3,...,k\})$
\begin{equation}
p_r^{(k)}(\mu) = \frac{(-1)^{r-1}}{(r-1)!(k-r)!} \cdot \frac{1-\mu}
{(1-\mu)(r-1)+1} \cdot \left(\frac{2-\mu}{1-\mu}\right)_{k-1}.
\label{48}
\end{equation}

In the same context we can calculate the critical value for
\begin{eqnarray}
b(0)& = & \frac{2\Pi_2r_0}{\Pi_1} = - a(0)^{-1} \nonumber \\
& = & - \left(\frac{\Pi^2_1}{2\Pi_2}\right)^{-1} \Pi_1 U^{\prime}(\Pi_1)
\label{49}
\end{eqnarray}
which gives
\begin{eqnarray}
b(0) &=& p_1^{(k)}(\mu) \nonumber \\
&=& (1-\mu) \left[\frac{1}{(k-1)!} \left(\frac{2-\mu}{1-\mu}\right)_{k-1}-1
\right]
\label{50}
\end{eqnarray}
(if $r=1$ is inserted into (\ref{48}) we obtain $p_1^{(k)}(\mu)+(1-\mu)$).

At the critical point also $g_{k+1}$ (\ref{34}) is fixed
\begin{equation}
\frac{g^c_{k+1}}{k+1} = (-1)^{k+1} \frac{(p_1^{(k)}(\mu))^{k+1}}{(k+1)!}
\left(\frac{2-\mu}{1-\mu}\right)_{k-1} \frac{\Pi^2_1}{2\Pi_2}.
\label{51}
\end{equation}

It is important to remark that the second condition (\ref{34}) is satisfied
indeed by (\ref{51}), since for $0 \le \mu < 1$ $(0 \le D < 2)$
\begin{equation}
\left(\frac{2-\mu}{1-\mu}\right)_{k-1} > (k-1)!
\label{52}
\end{equation}
\begin{equation}
p_1^{(k)}(\mu) > 0
\label{53}
\end{equation}
\begin{equation}
\mbox{sign }g^c_{k+1} = (-1)^{k+1}.
\label{54}
\end{equation}

\section{Deformation of the singularity and the double scaling limit for
fixed finite volume}

Singularities can be deformed \cite{11,12}. We assume that this is achieved for
$A_k$ by
\begin{equation}
f_r = f^c_r(1+\Theta_r)
\label{55}
\end{equation}
\[
2 \le r \le k
\]
whereas the parameter $m^2$ is invariant. This is a nonstandard way of
deforming: the standard way would be to keep $f_k = f^c_k$ fixed and let $m^2$
(or $f_1$) vary. The quantity $m^{2}$ is kept constant to simplify the
following
discussion and this is achieved by compensation of the variation of $r_0$ and
$\beta^{2}$ in (\ref{8})
\begin{equation}
m^{2} = \beta^{2} - 2 r_0.
\end{equation}
Invariance of $m^{2}$ implies that
\begin{equation}
\sigma_{0} = \Pi_{1}(m^{2})
\end{equation}
is not varied either, so that in
\begin{equation}
r_0 = -U'(\sigma_{0})
\end{equation}
the variation comes only from (\ref{55}). We conclude that from the $k$
quantities
$\{ f_1=\frac{\beta^{2}}{2}, f_2, \ldots, f_k \}$ only $k-1$ are varied.

In any case the Hessian is diagonalized exactly and the fluctuations ${\hat
\xi},
{\hat \eta}$ are considered independent of the deformation.

Correspondingly the coupling constants $\{g_n\}$ in (\ref{33}) deviate from
the critical values,
\begin{equation}
\frac{g_n}{n} = \sum^k_{r=2} \alpha_{nr}^{(k)} \Theta_r + O_2(\Theta)
\label{56}
\end{equation}
\[
(2 \le n \le k) .
\]
The double scaling limit is obtained by coupling two processes

\quad (1)\quad$N \rightarrow \infty$

\quad (2)\quad$\Theta_r \rightarrow 0, \: \forall r$

\noindent in a particular way. In the momentum spectrum any
neighboring eigenvalue $p_1$ of $p = 0$ has a fixed distance $O(L^{-1})$ of
$p = 0$ and therefore the deformation parameters can be restricted to a
neighborhood of zero so small that $\lambda_-(p_1)$ and $\lambda_-(0)$ have
values in non-intersecting intervals on this neighborhood.

We study the singular partition function
\begin{equation}
Z_{\mbox{\scriptsize sing}} = \int d\eta_0 e^{-NS_{\mbox{\tiny red}}(\eta_0)}.
\label{57}
\end{equation}
Since $g^c_{k+1} \neq 0$ we can approximate $g_{k+1}$ by $g^c_{k+1}$ over the
whole deformation neigborhood. We introduce a new variable $y$ by
requiring

\begin{equation}
y^{k+1} = N |g^c_{k+1}| \eta_0^{k+1}
\label{58}
\end{equation}

\begin{equation}
\eta_0 = y\left(N|g^c_{k+1}|\right)^{-\frac{1}{k+1}}.
\label{59}
\end{equation}
Inserting (\ref{59}) into $S_{\mbox{\scriptsize red}}(\eta_0)$, we see that we
have to perform
the limit \newpage

\begin{equation}
\zeta_n = \lim N \left(N|g^c_{k+1}|\right)^{-\frac{n}{k+1}}
\sum_r \alpha^{(k)}_{nr} \Theta_r
\label{60}
\end{equation}
\[
(2 \le n \le k).
\]
Thus the double scaling limit is defined via the "susceptibility matrix"
$\alpha^{(k)}$, and each linear combination $(\alpha^{(k)} \Theta)_{n}$ scales
as $N^{-\sigma_n^{(k)}}$
\begin{equation}
\sigma_n^{(k)} = 1 - \frac{n}{k+1}.
\label{61}
\end{equation}
These are the "critical indices". The result of the double scaling limit is
to leading order
\begin{equation}
Z_{\mbox{\scriptsize sing}} = \left(N|g^c_{k+1}|\right)^{-\frac{1}{k+1}}
Y_\epsilon(\zeta_2,
\zeta_3,...\zeta_k)
\label{62}
\end{equation}
where $Y_\epsilon$ is the generalized Airy function,
\begin{equation}
Y_\epsilon(\zeta_2,\zeta_3,...,\zeta_k) = \int_{C^{(k)}} dy \exp
\left\{-\sum^k_{n=2} \zeta_n y^n - \epsilon \frac{y^{k+1}}{k+1} \right\}
\label{63}
\end{equation}
\begin{equation}
\epsilon = \mbox{sign }g^c_{k+1}.
\label{64}
\end{equation}
The contour $C^{(k)}$ is the real axis if
\begin{equation}
\epsilon = + 1, \; k \mbox{ odd}
\label{65}
\end{equation}
and a combination of complex contours, running from infinity to infinity along
which the integral converges exponentially, in all other cases. By a
translation
$y \rightarrow y+a$ we can eliminate the term $y^k$ and produce a term $y^1$
obtaining the standard form of a generalized Airy function.

The function $Y_{\epsilon}$ or any function of $Y_{\epsilon}$ such as
$F = \log Y_{\epsilon}$ satisfy a renormalization group equation
\begin{equation}
\left( N \frac{\partial}{\partial N} - \sum^k_{n=2} \beta_n(\Theta) \frac
{\partial}{\partial\Theta_n}\right) F(\zeta_2,\zeta_3,...\zeta_k) = 0
\label{66}
\end{equation}
where each $\zeta_n$ is considered as a function of $N$ and $\{\Theta_r\}$
which in the neighborhood of the singularity is determined by (\ref{60}). The
beta functions $\{\beta_n(\Theta)\}$ are determined from $(2 \le n\le k)$
\begin{equation}
N \frac{\partial\zeta_n}{\partial N} - \sum^k_{r=2} \beta_r(\Theta) \frac
{\partial\zeta_n}{\partial\Theta_r} = 0.
\label{67}
\end{equation}
For small $\{\Theta_r\}$ this is satisfied if
\begin{equation}
\beta_r(\Theta) = \sum^k_{n=2} {\cal N}^{(k)}_{rn} \Theta_n + O_2(\Theta)
\label{68}
\end{equation}
\begin{equation}
{\cal N}^{(k)} = \alpha^{(k),-1} \mbox{diag }\sigma^{(k)} \alpha^{(k)}.
\label{69}
\end{equation}

The susceptibility matrix must therefore be invertible. We will verify this by
an explicit calculation. This calculation starts from the following
observation. We can use
\begin{equation}
\left.u_n = U^{(n)}(\Pi_1) - U^{(n)}(\Pi_1)\right|_{crit}
\label{70}
\end{equation}
as parameters of deformation instead of the $\{\Theta_n\}$. Then $\tilde{S}
_{\mbox{\scriptsize red}}$ depends on
\begin{equation}
\tilde{S}_{\mbox{\scriptsize red}} (\xi_0,\eta_0;u_2,u_3,...u_k).
\label{71}
\end{equation}
{}From the constraint (\ref{30}) we obtain the elimination function
\begin{equation}
\xi_0 = H(\eta_0;u_2,u_3,...,u_k)
\label{72}
\end{equation}
and
\begin{equation}
S_{\mbox{\scriptsize red}}(\eta_0;u_2,u_3,...,u_k) =
\tilde{S}_{\mbox{\scriptsize red}}(H(\eta_0;u_2,u_3,...,u_k),
\eta_0;u_2...u_k).
\label{73}
\end{equation}
It follows from (\ref{30}) that in
\begin{equation}
\frac{\partial S_{\mbox{\scriptsize red}}}{\partial u_n} =
\frac{\partial\tilde{S}_{\mbox{\scriptsize red}}}
{\partial \xi_0} \frac{\partial H}{\partial u_n} + \frac{\partial
\tilde{S}_{\mbox{\scriptsize red}}}
{\partial u_n}
\label{74}
\end{equation}
the first term vanishes. The variation of $u_n$ enters the second term either
directly
or via $r_0$. However, it can be shown, that at the critical point and for
constant
$\{ a_n \}$ the derivative with respect to $r_0$ vanishes. Thus only the direct
dependence
is left and we obtain
\begin{eqnarray}
\frac1n \frac{\partial g_n}{\partial u_{\ell}} = \frac{1}{\ell !} \left(- \frac
{\Pi_1}{b(0)}\right)^{\ell} \sum_{{\mbox{\scriptsize partitions of n}} \atop
{\mbox{\scriptsize of length $\ell$}}}
\left(\begin{array}{cc}
\ell \\
n_1n_2n_3...
\end{array}\right) \nonumber \\
(-b(0)^2)^n \prod^{\infty}_{j=2} a_j^{n_j}
\label{75}
\end{eqnarray}
where $\ell$ is the length and $n_j$ the repetition number of $j$ (as in
(\ref{42}),(\ref{43})). The coefficients $\{a_j\}$ of the elimination
function $H$ (\ref{32}) at the critical point can be expressed as functions
of $D$ (or $\mu$) by
\begin{eqnarray}
a_{n+1} = - b(0)^{n+2} \sum_{\mbox{\scriptsize partitions of $n$}}
\frac{(n+\ell)!}
{(n+1)!} (1+b(0)^2)^{-\ell} \nonumber \\
\cdot \prod^{\infty}_{j=1} \frac{1}{n_j!} \left(\frac{v_{j+2}}{(j+1)!}
\right)^{n_j}
\label{76}
\end{eqnarray}
with $\ell$ and $n_j$ as in (\ref{42}), (\ref{43}). This formula has been
verified for $n \le 8$.

Next we reduce the susceptibility matrix
\begin{equation}
\alpha_{nr}^{(k)} = \frac{\Pi^2_1}{2\Pi_2} (p_1^{(k)}(\mu))^n p_r^{(k)}(\mu)
\tilde{\alpha}^{(k)}_{nr}\label{77}
\end{equation}
and find from inserting (\ref{76}) into ${\tilde S}_{\mbox{\scriptsize red}}$
\begin{equation}
\alpha_{nr}^{(k)} = \sum^r_{s=2} S^{(k)}_{ns} \left(r-1 \atop{s-1}\right).
\label{78}
\end{equation}

Here $S^{(k)}$ is a lower left triangular matrix
\begin{equation}
S^{(k)}_{nr} = 0,  \quad r > n
\label{79}
\end{equation}
and
\begin{equation}
B_{sr} = \left(r-1 \atop{s-1}\right)
\label{80}
\end{equation}
is an upper right triangular matrix. Moreover we have
\begin{equation}
S^{(k)}_{nn} = \frac1n \quad (n \geq 2)
\label{81}
\end{equation}
and for all other elements $n > r \geq 2$
\begin{eqnarray}
S^{(k)}_{nr} = \sum_{\mbox{\scriptsize partitions of $n-r$}} (1
-\frac{n}{n+\ell-1} \delta_{r2})
\frac{(n+\ell-1)!}{n!} (1+b(0)^2)^{-\ell} \nonumber \\
\cdot \prod^{\infty}_{j=1} \frac{1}{n_j!} \left(\frac{v_{j+2}}{(j+1)!}\right)
^{n_j}
\label{83}
\end{eqnarray}
with $\ell$ the length and $n_j$ the repetition number of $j$ of the partition
of
$n-r$.

The representation (\ref{78}) gives for the inverse

\begin{equation}
\tilde{\alpha}^{(k),-1} = B^{-1} S^{(k),-1}
\label{84}
\end{equation}
with
\begin{equation}
B^{-1}_{sr} = (-1)^{s-r} \left(r-1 \atop{s-1} \right)
\label{85}
\end{equation}
and
\begin{eqnarray}
S^{(k),-1}_{rn} &=& - rn S^{(k)}_{rn} + \sum_{s \atop{(r>s>n)}} rsn
S^{(k)}_{rs} S^{(k)}_{sn} \nonumber \\
& & - \sum_{s_1,s_2 \atop{(r>s_1>s_2>n)}} rs_1s_2n
S^{(k)}_{rs_1}S^{(k)}_{s_1s_2}
S^{(k)}_{s_2n} \nonumber \\
& & \mp \ldots \quad (r > n)
\label{86}
\end{eqnarray}
\begin{equation}
S^{(k),-1}_{nn} = n.
\label{87}
\end{equation}
Finally we obtain by inserting (\ref{77}),(\ref{78}), and (\ref{84}) into
(\ref{69})
\begin{equation}
{\cal N}^{(k)}_{rn} = \frac{p_n^{(k)}(\mu)}{p_r^{(k)}(\mu)} \left( B^{-1}
S^{(k),-1} \mbox{diag } \sigma^{(k)}S^{(k)}B\right)_{rn}.
\label{88}
\end{equation}

\section{The case of unbounded volume}

In the case of unbounded volume the momentum spectrum is continuous. The
domain of small momenta
\begin{equation}
|p| < \Lambda
\label{89}
\end{equation}
is considered as an additional deformation. This leads to an additional
term in the generalized Airy function integral (the kinetic energy term) and
modified critical indices
\begin{equation}
\sigma_n^{(k)} \rightarrow \chi_n^{(k)}(\mu).
\label{90}
\end{equation}
The generalized Airy function (\ref{63}) is replaced by a field theoretic
partition function
\begin{eqnarray}
Y_{\phi} = \int D\phi \exp \left\{- \frac12 \int d^Dx \phi(x)(-\Delta+M^2)
\phi(x) \right. \nonumber \\
\left. - \sum^k_{n=3} \zeta_k \int d^Dx \phi(x)^k - \frac{F_{k+1}}{k+1} \int
d^Dx \phi(x)^{k+1} \right\}.
\label{91}
\end{eqnarray}
The kinetic energy term has been normalized in (\ref{91}) instead of the
$(k+1)$-st order term. The dimension $D$ is still in the interval
$0 \leq D <2$.

The reduced action $S_{\mbox{\scriptsize red}}$ depends only on
$\{\hat{\eta(p)} \, | \, |p|<\Lambda\}$
so that all other degrees of freedom must be integrated out by performing
a Gaussian saddle point integration. As the first step we obtain the
half-reduced
action
\begin{equation}
\left.\tilde{S}_{\mbox{\scriptsize red}}(\hat{\xi}(p),\hat{\eta}(p)) =
S_{\mbox{\scriptsize eff}}(\hat{\xi},
\hat{\eta})\right|_{\hat{\xi}(p)=\hat{\eta}(p)=0 \atop{|p|>\Lambda}}.
\label{92}
\end{equation}
We are left with the issue to solve
\begin{equation}
\frac{\delta}{\delta\hat{\xi}(p)} \tilde{S}_{\mbox{\scriptsize red}} = 0
\mbox{ for all } |p| < \Lambda
\label{93}
\end{equation}
for $\hat{\xi}(p)$. If $\Lambda \ll m$ the trace terms
\begin{equation}
\mbox{Tr}[(-\Delta+m^2)^{-1} \beta(x)]^n = \int \prod^n_{i=1} \frac{d^Dq_i}
{(2\pi)^D} \frac{\hat{\beta}(q_i-q_{i+1})}{q^2_i+m^2}
\label{94}
\end{equation}
\[
(q_{n+1} = q_1)
\]
reduce to the n-fold convolution product of $\hat{\beta}$ at argument zero
times a constant (see (\ref{36})), namely
\begin{equation}
= \Pi_n \hat{\beta}^n_{\ast}(0).
\label{95}
\end{equation}
Thus the elimination (\ref{93}) differs from (\ref{30}), (\ref{31}) only
by the replacement of $\xi_0^{n_1} \eta_0^{n_2}$ by
\begin{equation}
\left(\hat{\xi}_{\ast}^{n_1} \ast \hat{\eta}_{\ast}^{n_2}\right)(0).
\label{96}
\end{equation}
Its solution is
\begin{equation}
\hat{\xi}(p) = \sum^{\infty}_{n=2} a_n \hat{\eta}^n_{\ast}(p)
\label{97}
\end{equation}
with $\{a_n\}$ as in (\ref{32}). Correspondingly the reduced action is
\begin{equation}
S_{\mbox{\scriptsize red}}(\hat{\eta}) = \sum^{\infty}_{n=2} \frac{g_n}{n}
\hat{\eta}^n_{\ast}(0)
\label{98}
\end{equation}
with $\{g_n\}$ as in (\ref{33}). The condition (\ref{34}) for the appearance of
a
singularity $A_k$ remains unchanged.

Momenta and coordinates are scaled by \cite{8,9}
\begin{equation}
p = N^{-\lambda}p^{\prime}
\label{99}
\end{equation}
\begin{equation}
x = N^{+\lambda}x^{\prime}
\label{100}
\end{equation}
where $\lambda >0$ is necessary in order that in the limit $N \rightarrow
\infty$
the domain $\Lambda$ is mapped onto $\bbbr_D$. The fields are renormalized by
\begin{equation}
\phi(x^{\prime}) = C^{(k)} N^{\frac{1+D\lambda}{k+1}} \eta(x)
\label{101}
\end{equation}
so that the power of order $k+1$ in (\ref{91}) obtains a finite coefficient
in the limit $N \rightarrow \infty$. Both $C^{(k)}$ and $\lambda$ are
determined from the kinetic energy term.

We return to
\begin{equation}
\frac12 \int\limits_{|p|<\Lambda} \frac{d^Dp}{(2\pi)^D} \lambda_-(p) N_-(p)
\hat{\eta}(-p)\hat{\eta}(p)
\label{102}
\end{equation}
and expand into deformation parameters $\{\Theta_n\}$ and $p$
\begin{eqnarray}
\lambda_-(p)N_-(0) &=& \frac{\Pi^2_1}{2\Pi_2} b(0)^2 \Big\{\frac16(2-\mu)
\frac{p^2}{m^2} \nonumber \\
& &+ \sum^k_{r=2} (r-1) p_r^{(k)}(\mu)\Theta_r \Big\} \mbox{ + higher
order terms}
\label{103}
\end{eqnarray}
where (\ref{19}), (\ref{52}), (\ref{46}) have been used. This implies
(see (\ref{50}))
\begin{equation}
C^{(k)} = p_1^{(k)} (\mu) \left[ \frac{\Pi^2_1}{2\Pi_2} \frac16 (2-\mu)
\frac{1}{m^2}\right]^\frac12
\label{104}
\end{equation}
and \cite{9}
\begin{equation}
\lambda = \frac{k-1}{2(k+1)-D(k-1)}.
\label{105}
\end{equation}
Positivity of $\lambda$ is satisfied as long as
\begin{equation}
D < D_{\infty} = 2 \frac{k+1}{k-1}.
\label{106}
\end{equation}
This is trivially fulfilled for $D < 2$. From the second term in (\ref{100})
we obtain the double scaling limit
\begin{equation}
\lim_{N \rightarrow \infty \atop{\Theta_r \rightarrow 0, \mbox{\scriptsize
{ all r}}}} N^{2\lambda} \left(\sum^k_{r=2}(r-1) p_r^{(k)}(\mu)\Theta_r\right)
=
\frac16 (2-\mu) \frac{M^2}{m^2}.
\label{107}
\end{equation}
Since this limiting procedure is independent of the other double scaling
limits described below (due to the invertibility of the susceptibility matrix)
we can ascribe to $M^2$ any value, in particular any positive value.

The susceptibility matrix $\alpha^{(k)}$ (\ref{53}) enters all other
double scaling limits as usual. For all $3 \le n \le k$ we have
\begin{equation}
\zeta_n = \lim_{N \rightarrow \infty \atop{\Theta_r \rightarrow 0,
\mbox{\scriptsize
{ all r}}}} (C^{(k)})^{-n} N^{\chi_n^{(k)}} \sum^k_{r=2} \alpha^{(k)}_{nr}
\Theta_r
\label{108}
\end{equation}
where the critical indices are now
\begin{equation}
\chi_n^{(k)} = \frac{k+1-n}{(k+1)-\mu(k-1)}
\label{109}
\end{equation}
so that
\begin{equation}
\chi_n^{(k)}(\mu=0) = \sigma_n^{(k)}
\label{110}
\end{equation}
(see (\ref{61})). Moreover we find from (\ref{109})
\begin{equation}
\chi_2^{(k)} = 2\lambda.
\label{111}
\end{equation}
If we identify
\begin{equation}
\zeta_2 = \frac12 M^2
\label{112}
\end{equation}
we can incorporate the limit (\ref{107}) into the set of limits (\ref{108})
as in the case $n=2$.

Finally we note that (see (\ref{51}), (\ref{101}))
\begin{equation}
F_{k+1} = (C^{(k)})^{-k-1} g^c_{k+1}.
\label{113}
\end{equation}
The partition function is written as function
\[
Y_{\phi}(\zeta_2,\zeta_3,...,\zeta_k)
\]
of the double scale invariant quantities $\{\zeta_n\}$. Any function of
$Y_{\phi}$ satisfies a renormalization group equation (\ref{66}) with
beta functions obeying (\ref{68}), where, however,
\begin{equation}
{\cal N}^{(k)} = \alpha^{(k),-1} \mbox{ diag} \chi^{(k)} \alpha^{(k)}.
\label{114}
\end{equation}

\section{Dimensions $D \ge 2$}

At $D=2$ the integral $\Pi_1$ (\ref{36}) exhibits a pole of first order
whereas $\Pi_n, n \ge 2$, are holomorphic
\begin{equation}
\Pi_1 = \frac{1}{4\pi} \frac{1}{1-\mu} + \frac{1}{4\pi} \log
\frac{4\pi e^{\Gamma^{\prime}(1)}}{m^2} + O(1-\mu).
\label{115}
\end{equation}
To obtain a regular expression in $2<D<4$ we can simply analytically continue
$\Pi_1$ in $D$:
\begin{equation}
\Pi_1^{\mbox{\scriptsize an}} = \int \frac{d^Dp}{(2\pi)^D}
\left[(p^2+m^2)^{-1}-(p^2)^{-1}\right]
\label{116}
\end{equation}
is the convergent integral representation for this analytic continuation
in $2<D<4$. We can thus renormalize the position of the saddle point
$\sigma_0 \rightarrow \sigma_0^{\mbox{\scriptsize ren}}$
\begin{equation}
\sigma_0^{\mbox{\scriptsize ren}} = \Pi_1^{\mbox{\scriptsize an}}.
\label{117}
\end{equation}
The purely formal subtraction formula
\begin{equation}
\sigma_0^{\mbox{\scriptsize ren}} = \sigma_0 - \sigma_{\infty}
\label{118}
\end{equation}
\begin{equation}
\sigma_{\infty} = \int \frac{d^Dp}{(2\pi)^D} \cdot \frac{1}{p^2} \qquad
{\mbox{(divergent)}}
\label{119}
\end{equation}
suggest how to renormalize coupling constants and mass \cite{17}. We set
\begin{eqnarray}
U^{\prime}(\sigma_0) &=& \sum^{\infty}_{r=2} f_r\sigma_0^{r-1} \nonumber \\
&=& f_1^{\mbox{\scriptsize ren}} + \sum^{\infty}_{r=2} f_r^{\mbox{\scriptsize
ren}}(\sigma_0^{\mbox{\scriptsize ren}})^{r-1}
\label{120}
\end{eqnarray}
where
\begin{equation}
f_n^{\mbox{\scriptsize ren}} = \sum^{\infty}_{r=n} \left(r-1 \atop{n-1}\right)
\sigma^{r-n}
_{\infty} f_r
\label{121}
\end{equation}
and $f_n^{{\mbox{\scriptsize ren}}}, 2 \le n \le k$, are assumed to be finite.
This can be
achieved
when (\ref{45}) is valid by adjusting $\{f_r \, | \, 2 \le r\le k\}$
correspondingly.
We renormalize $\rho_0$ by
\begin{equation}
i\rho^{{\mbox{\scriptsize ren}}} = i\rho_0 - f_1^{{\mbox{\scriptsize ren}}}
\label{122}
\end{equation}
and
\begin{equation}
(m^{{\mbox{\scriptsize ren}}})^2 = m^2-2f_1^{{\mbox{\scriptsize ren}}}.
\label{123}
\end{equation}
Finally we skip the label "ren" (and "an") and end up with the rule:
replace $\sigma_0$ by $\Pi_1^{{\mbox{\scriptsize an}}}$ and obtain all critical
quantities
in the interval $2<D<4$ by analytic continuation. The consistency of this
rule has still to be investigated.

First we inspect from (\ref{103}) that the sign of the kinetic energy term
remains unchanged and $C^{(k)}$ (\ref{104}) stays real. There remains,
however, a problem with the sign of
\begin{equation}
\lambda = \frac12 \chi_2^{(k)}
\label{124}
\end{equation}
((\ref{105}), (\ref{109}), (\ref{112})). We mentioned already that $\lambda$
is positive if $D<D_{\infty}(k)$. On the other hand the field theory
(\ref{91}) is superrenormalizable for $D<D_{\infty}(k)$ and renormalizable
for $D=D_{\infty}(k)$. Thus $D_{\infty}(k)$ is an absolute limit which we
cannot overcome.

In the case $D<D_{\infty}(k)$ we have to subtract some low order Green
functions,
for $D=D_{\infty}(k)$ we need a finite number of subtractions (counter terms)
in the action. The subtractions are by momentum cutoff
\begin{equation}
|p^{\prime}| \le N^{\lambda} \Lambda, \quad D < D_{\infty}(k).
\label{125}
\end{equation}
For $D=D_{\infty}(k)$ the limit $N \rightarrow \infty$ cannot be performed
at all but $N$ has to be renormalized
\begin{equation}
N^{\prime} = N^{\frac{1}{2(k+1)-D(k-1)}}, \quad D < D_{\infty}(k)
\label{126}
\end{equation}
\begin{equation}
|p^{\prime}| \le (N^{\prime})^{k-1} \Lambda
\label{127}
\end{equation}
and then the limit $N^{\prime} \rightarrow \infty, D \rightarrow D_{\infty}$
is executed. In any way we conclude that:
\begin{enumerate}
\item the parameter $\Lambda$ introduced in the scaling (\ref{99}), (\ref{100})
transforms the $N \rightarrow \infty$ limit to the UV-cutoff removal limit;
\item the double scaling limit makes sense only if the necessary subtractions
dictated by standard renormalization theory are performed in the limiting
procedure.
\end{enumerate}
New insights on the renormalization procedure in general cannot be expected.

For $D=2$ we replace analytic regularization by subtraction of the pole
term, i.e. from (\ref{115}) we derive
\begin{equation}
\Pi_1^{{\mbox{\scriptsize ren}}} = \frac{1}{4\pi} \log \frac{\mu^2}{m^2}.
\label{128}
\end{equation}
Identifying $\sigma_{\infty}$ in (\ref{118}) with the pole term (\ref{115})
we can proceed exactly as for $D > 2$ and renormalize coupling constants and
mass. The calculation of the critical quantities has to be redone in view of
(\ref{128}) and we will outline the results.

For $n \geq 2$ we obtain
\begin{equation}
\Pi_n = \frac{(m^2)^{1-n}}{4\pi(n-1)}.
\label{129}
\end{equation}
Inserting this into an equivalent version of (\ref{41}) we obtain
\begin{equation}
U^{(n)}(\Pi_1) = \frac12 m^2 (-4\pi)^{n-1}
\label{130}
\end{equation}
and correspondingly from (\ref{40}), (\ref{129})
\begin{equation}
v_n = - \left(- \log \frac{\mu^2}{m^2}\right)^{n-2}.
\label{131}
\end{equation}
The critical coupling constants follow from (\ref{130})
\begin{equation}
f^c_n = \frac12 m^2 \frac{(-4\pi)^{n-1}}{(n-1)!} j_{k-n} (\exp; \log
\frac{\mu^2}{m^2})
\label{132}
\end{equation}
where
\begin{equation}
j_n(f;z)
\label{133}
\end{equation}
is the Taylor polynomial ("jet") of degree n for the function f with
variable $z$. From (\ref{132}) we deduce
\begin{equation}
b(0) = \frac{j_{k-1}(\exp; \log \frac{\mu^2}{m^2})-1}{\log\frac{\mu^2}{m^2}}
\label{134}
\end{equation}
and
\begin{equation}
\frac{g^c_{k+1}}{k+1} = \frac{m^2}{8\pi} \frac{(1-j_{k-1}(\exp; \log
\frac{\mu^2}{m^2}))
^{k+1}}{(k+1)!}.
\label{135}
\end{equation}
The coefficients $\{a_n\}$ of the elimination function $H$ (\ref{32}) at the
critical
point are obtained from (\ref{76}) and (\ref{131})
\begin{eqnarray}
a_{n+1} &=& (-1)^{n+1} \frac{b(0)^{n+2}}{(n+1)!} (\log \frac{\mu^2}{m^2})^n
\nonumber \\
& & \cdot \sum^n_{\ell=1} (-1)^{\ell} \frac{K_{n\ell}}{(1+b(0)^2)^{\ell}}
\label{136}
\end{eqnarray}
where $K_{n\ell}$ are integers
\begin{equation} \label{137}
K_{n\ell} = (n+\ell)! \sum_{{\mbox{\scriptsize partitions of $n$}} \atop
{\mbox{\scriptsize of length $\ell$}}} \; \prod_{j=1}^{\infty}
\frac{1}{{n_j}!} \left( \frac{1}{(j+1)!} \right)^{n_j}
\end{equation}
($n_j$ is the repetition number of j in the partition).
Instead of (\ref{77}) we reduce the susceptibility matrix by
\begin{equation} \label{138}
\alpha_{nr}^{(k)} = (\Pi_1 b(0))^{n} (4 \pi)^{r} f_{r}^{c}
{\tilde{\alpha}}_{nr}^{(k)}
\end{equation}
and for the reduced matrix (\ref{78})-(\ref{83}) remain valid. Instead of
(\ref{83}) we find an analogous formula with $v_{j+2}$ replaced by $(-1)^{j+1}$
leading to
\begin{equation} \label{139}
S_{nr}^{(k)} = \sum_{\ell=1}^{n-r} (-1)^{n+\ell-r} (1-\frac{n}{n+\ell-1}
\delta_{r2}) \frac{(n+\ell-1)!}{n!}
\frac{K_{n-r,\ell}}{(n+\ell-r)!} (1+b(0)^{2})^{-\ell}.
\end{equation}

Since the double scaling limit for $D \geq 2$
necessitates regularization in the UV momentum domain, as dictated by the known
renormalization theory, we refrain from dealing with the case $D=4$ $(k=3)$
here.
It has been argued that a double scaling limit does not exist at $D=4$
\cite{18,19}.
We emphasize that $D=4$ is an "exceptional dimension" in the sense described
below
($n=2$ in (\ref{142}), $k_{\mbox{\scriptsize max}}=0$ in (\ref{144})). This
hints also to the
nonexistence of the standard double scaling limit.

Now we return to the second condition (\ref{34}) in the case $2 \leq D <
\infty$.
In fact from (\ref{51}) we can see that $g_{k+1}^{c}$ vanishes if and only if
\begin{enumerate}
\item $\left( \frac{2-\mu}{1-\mu} \right)_{k-1} = 0$, which is fulfilled for
\begin{equation} \label{140}
\mu = \frac{n}{n-1}, \quad 2 \leq n \leq k
\end{equation}
\item $p_{1}^{(k)}(\mu) = 0$, which occurs only if $k$ is odd and
\begin{equation} \label{141}
\mu = \frac{k+1}{k}
\end{equation}
\end{enumerate}
(that there are no other zeros of $p_{1}^{(k)}(\mu)$ in the interval $1 < \mu <
2$
has been verified by computer up to $k=20$).

Thus there exist exceptional dimensions
\begin{equation} \label{142}
\mu_n = \frac{n}{n-1}, \quad n \geq 3 \; (\in \bbbz)
\end{equation}
for which the type of the singularity $A_k$ has $k$ not constrained by the
renormalizability limit (see (\ref{107}))
\begin{equation} \label{143}
k \leq k_{\mbox{\scriptsize ren}} = \frac{\mu+1}{\mu-1}
\end{equation}
but by the stronger bound
\begin{equation} \label{144}
k \leq k_{\mbox{\scriptsize max}} = \left\{ {{n-1, \quad n \mbox{ odd,}} \atop
{n-2, \quad n \mbox{ odd.}}} \right.
\end{equation}
In particular the physically very interesting case $D=3$ is in this set of
exceptional dimensions with $n=3$ and
\begin{equation} \label{145}
k_{\mbox{\scriptsize max}} = 2.
\end{equation}

\section{Remark: The unstable cases}

All actions (\ref{91}) for which
\begin{equation} \label{146}
\mbox{sign } F_{k+1}^{c} = +1, \quad k \mbox{ odd}
\end{equation}
is {\underline{not}} satisfied are unstable field theories if interpreted
conventionally. However, our derivation of these conditions from saddle point
integrals implies that the fields $\phi$ range over complex contours. For
example, for $k=2$, these are the standard Airy function contours
\begin{eqnarray} \label{147}
C &=& C_{0} -\frac{1}{2} (C_{\frac{1}{3}} + C_{\frac{2}{3}}) \qquad (\epsilon =
+1) \\
\label{148} C &=& -C_{\frac{1}{2}} +\frac{1}{2} (C_{\frac{1}{6}} +
C_{\frac{5}{6}})
\qquad (\epsilon = -1)
\end{eqnarray}
where $C_q, \; q \in \bbbq$, denotes the oriented ray along the argument
\begin{equation} \label{149}
\mbox{arg } C_q = 2 \pi q
\end{equation}
from zero to infinity. Again from $k=2, \; D=0$ we know that there exists a
domain of parameters
\begin{equation}
(\zeta_2, \zeta_3, \ldots, \zeta_k, F_{k+1})
\end{equation}
where the partition function is positive and another one where it oscillates
so that both domains are separated by a hypersurface on which the partition
function vanishes. Whether in the domain of positivity the action defines a
reasonable renormalizable field theory is unknown.

\end{document}